\documentclass[review,12pt,1p,times]{iopart} 
\usepackage{epsfig}

\begin{document}
\title{Refractive Index of p-SnS Thin Films and its Dependence on Defects}

\author{Yashika Gupta$^1$ $^2$, P.Arun$^2$}
\address{$^1$Department of Electronic Science, University of Delhi-South
Campus, New Delhi 110021, INDIA}
\address{$^2$Department of Electronics, S.G.T.B. Khalsa College, 
University of Delhi, Delhi 110007, INDIA}
\ead{arunp92@sgtbkhalsa.du.ac.in}

\begin{abstract}
Tin sulphide thin films of p-type conductivity were grown on glass substrates. 
The refractive index of the as grown films, calculated using both
Transmission and ellipsometry data were found to follow the Sellmeier dispersion
model. The improvement in the dispersion data obtained using ellipsometry
was validated by Wemple-Dedomenico (WDD) single oscillator model fitting.
The optical properties of the films were found to be closely related to the
structural properties of the films. The band-gap, its spread and appearance
of defect levels within the band-gap intimately controls the refractive
index of the films.
\end{abstract}


\section{Introduction}
Tin sulphide (SnS) is an earth-abundant, non-toxic, inorganic semiconductor 
material having layered structure. Its structural and optical properties can 
be easily varied according to the methods of fabrication~\cite{reddy1}-
\cite{nahass}. Literature shows that the SnS thin films can have both direct 
and indirect band-gaps ranging from 1.1-2.1~eV~\cite{gaoC, sohila}. Also, SnS 
can be of both p-type and n-type~\cite{ogah, leach}. All these properties 
make it a suitable as a photo-active layer in solar cells~\cite{gaoN,noguchi}, 
in near IR detectors~\cite{patel} and as an optical data storage 
media~\cite{yue1}. 

Considering all these applications are based on the optical/ 
refractive index properties of SnS films, this paper addresses itself on the 
variation of the optical properties of SnS thin films with film thickness.
Also, in our recent study on as grown SnS thin films (thickness between
450-960~nm), we observed persistent photocurrent (PPC) which decayed 
exponentially/ near exponentially with time~\cite{yppc}. We indicated that 
the variation in the time decay constant and in general the nature of decay 
might be associated with defect energy levels within the band-gap. The 
presence of defect energy levels (both acceptor and donor levels) within the 
band-gap is well documented in the literature~\cite{noguchi,juar}. 
Due to the large atomic size of Sulphur, one does not expect interstitial 
defects~\cite{wang}. Tin (Sn) vacancies are more predominant resulting in 
acceptor levels in SnS thin films~\cite{rist} while donor levels 
appear due to sulphur vacancies within the lattice~\cite{saj}. The present 
papers quest is not only to investigate the optical properties of p-SnS thin 
films of thicknesses 450-960~nm, but also to use the data to co-relate it with 
the PPC results of our previous studies.

\section{Experimental}
Tin sulphide thin films of varying thicknesses were grown at the rate of 
2~nm/sec on glass slides by thermal evaporation technique. Pellets of SnS 
powder (99\% purity) supplied by Himedia (Mumbai) were evaporated in vacuum 
better than ${\rm \approx 4\times 10^{-5}}$~Torr using a Hind High Vac (12A4D) 
thermal evaporation coating unit maintained at room temperature. Hotspot 
method confirmed the p-type conductivity of the as-grown films. Thickness of 
the films was measured using Vecco Dektak Surface Profiler (150) and the 
standard structural characterization of the films was done using Bruker D8 
X-ray diffractometer. UV-VIS double beam spectrophotometer (Systronics 2202) 
was used for optical characterization of the films. The surface morphology
of the samples was studied using a Field Emission-Scanning Electron
Microscope (FE-SEM FEI-Quanta 200F). Spectroscopic Ellipsometry (SE) studies 
of the samples were done using J.A. Woolam (USA) spectroscopic ellipsometer at 
an angle of incidence of ${\rm 65^o}$. Photoluminescence studies were done
using Shimadzu's spectrofluorophotometer Rf-5301PC at an excitation
wavelength of 550~nm in the wavelength range of 300-1000~nm. 

\section{Results and Discussion}
\subsection{The Structural Analysis}
\begin{figure}[h!!!]
\begin{center}
\epsfig{file=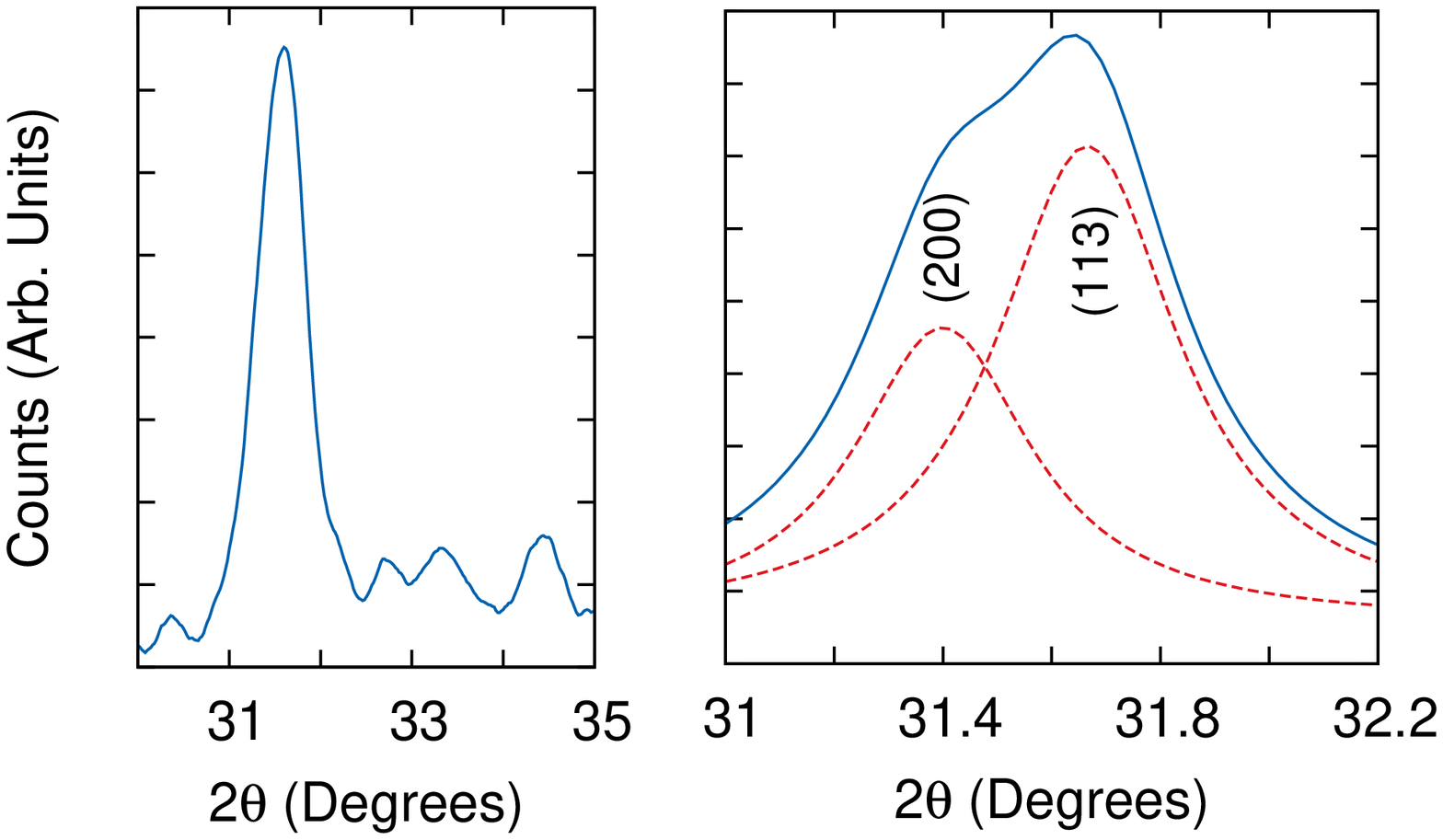, width=2.15in, angle=-90}
\end{center}
\caption{X-ray diffractogram of film with thickness 960~nm.
Plot showing deconvolution of broad peak at 
${\rm 2\theta \approx 31^{\circ}}$ indicating (113) as the preferred 
orientation for 960~nm thick film.}
\label{fig.yp7xrd}
\end{figure}

\begin{figure}[b!!!]
\begin{center}
\epsfig{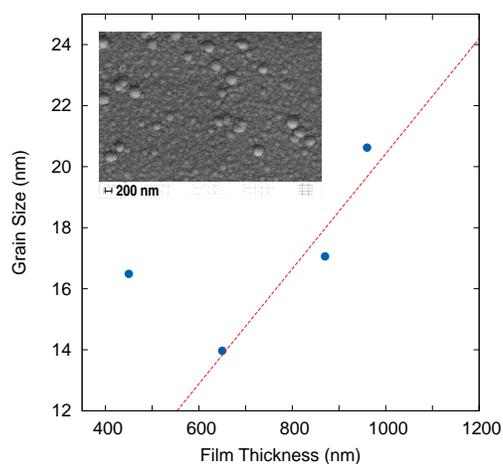}
\end{center}
\vskip -0.6cm
\caption{Plot shows variation in average grain size with thickness for the 
as-grown films. Inset shows spherical grains seen by Scanning Electron
Microscope. The micrograph exhibited here is of 870~nm thick SnS films.}
\label{fig.gsvst}
\end{figure}

Tin sulphide films in the thickness range of 450-1000~nm were used in this
study. The discussion on refractive index would justify the appropriateness
of the selection. X-ray diffraction studies of all the as-grown films 
without exception were nano-crystalline in nature with 450~nm thick sample
having an intense peak around ${\rm 2\theta \approx 38^{\circ}}$ while all 
other samples had a lone intense peak around 
${\rm 2\theta \approx 31.0^{\circ}}$ (fig~\ref{fig.yp7xrd}). 
Fig~(\ref{fig.yp7xrd}) is of a 960~nm thick film and is representative 
for all the samples. The peak positions (both intense and smaller peaks)
matched the peak positions as listed in ASTM Card No. 79-2193. This
suggests that our samples have an orthorhombic unit-cell structure with 
lattice parameters ${\rm a \approx 5.673~\AA}$, ${\rm b \approx 5.75~\AA}$ 
and ${\rm c \approx 11.76~\AA}$. The peak around 
${\rm 2\theta \approx 38^{\circ}}$ corresponds to the (005) plane while the 
peak at ${\rm 2\theta \approx 31.0^{\circ}}$ of the remaining samples was 
broad with a shoulder visible on enlarging the graph. This broadening of the 
peak was due to the very close, yet resolvable peaks, corresponding to the 
(113) plane (${\rm 2\theta \approx 31.788^{\circ}}$) and (200) plane 
(${\rm 2\theta \approx 31.515^{\circ}}$). 
Fig~(\ref{fig.yp7xrd}) also shows the deconvolution of this broad peak. The
deconvolution was done using standard software (Origin 6.0). Considering
that deconvolution gives us the true Full width at half maxima (FWHM) of the
Lorentzian XRD peaks, we can now calculate the average grain size of the 
as-grown films using the Scherrers formula \cite{cullity}
\begin{eqnarray}
D={ 0.9\lambda \over \beta cos \theta}\label{gs}
\end{eqnarray}
where `D' is the average grain size, ${\rm \lambda}$ is the wavelength of the 
X-ray used (${\rm \lambda=1.5406\AA}$), ${\rm \beta}$ is Full Width at Half 
Maximum intensity of the diffraction peaks and ${\rm \theta}$ is the Bragg's 
angle. The 0.9 coefficient is used for spherical grains, as was the case with
our samples (inset of Fig~\ref{fig.gsvst} shows spherical grains uniformly
dispersed on sample surface). Fig~(\ref{fig.gsvst}) shows a linear increase 
in average grain size proportional to the film thickness, suggesting an 
improvement in crystallinity of the films with a corresponding increase in 
the film thickness. However, the 450~nm thick film 
did not follow the trend. We believe this might be due to a different 
orientation of the planes compared to the other samples considering its very 
low film thickness.

\subsection{Optical Properties}
To study the optical properties of the films, absorption and transmission
spectra of the samples were analyzed in the wavelength range of 300-900~nm. 
The refractive index and band-gap of the films were calculated using the 
transmission and absorption data respectively. In the following passages we 
discuss our observations.

\subsubsection{Band-gap Variation}
\begin{figure}[h!!!]
\begin{center}
\epsfig{file=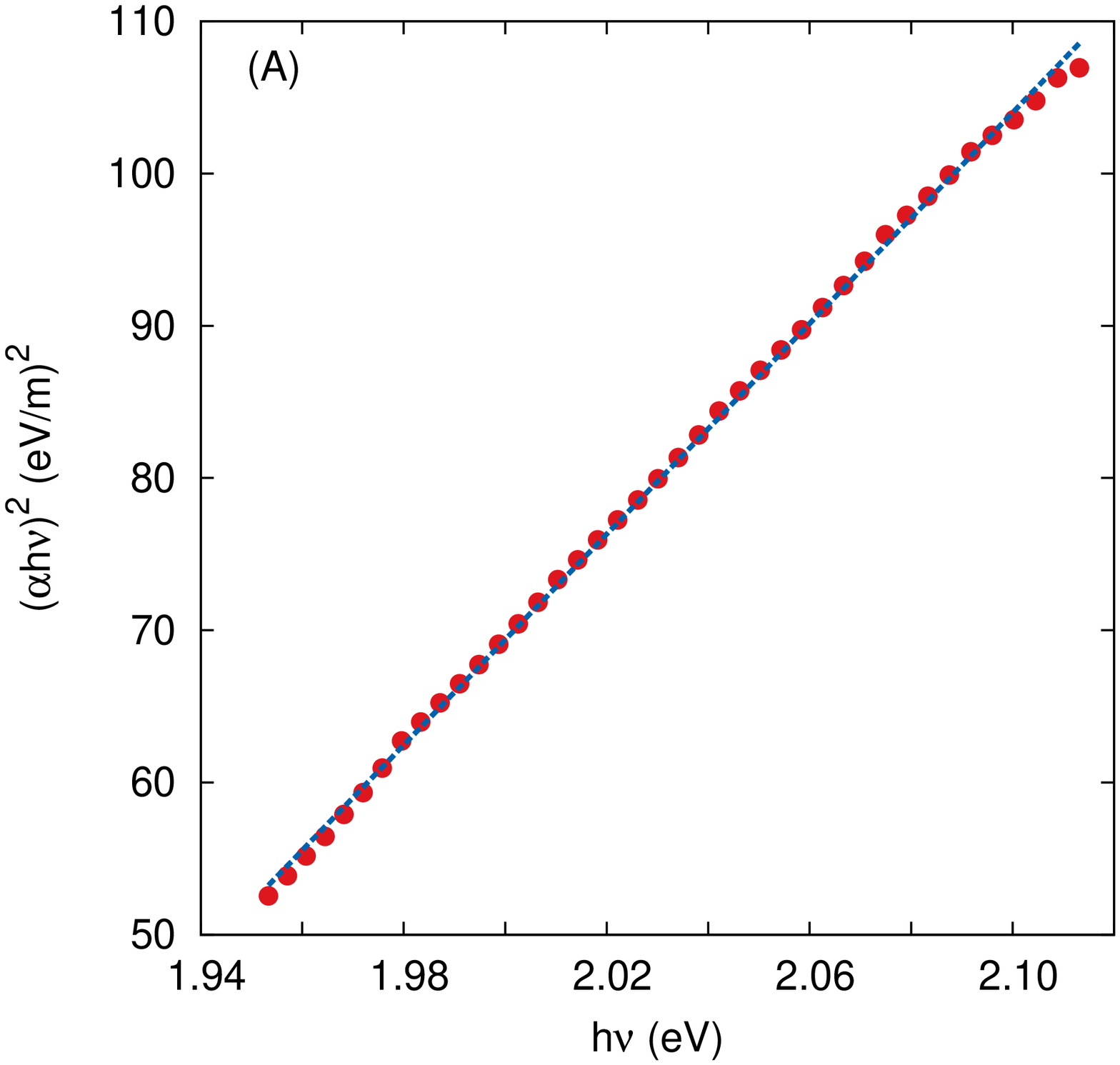, width=2.25in, angle=-90}
\hfil
\epsfig{file=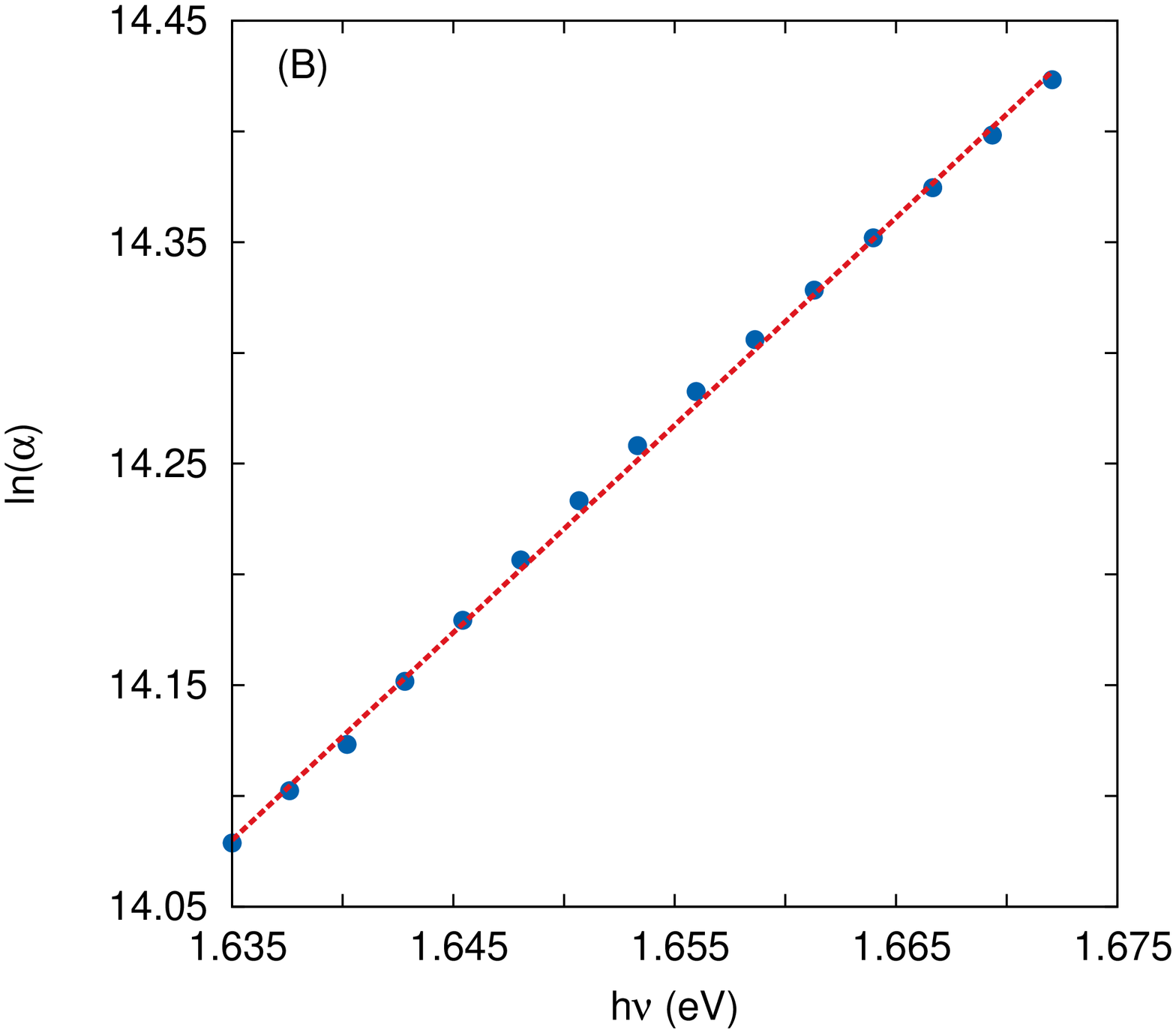, width=2.25in, angle=-90}
\end{center}
\caption{Plot exhibits variation of (A) ${\rm (\alpha h \nu)^2}$ with 
${\rm h\nu}$ for
650~nm thick film. Extrapolation of the best fit line to the X-axis at y=0 
gives the band gap of the samples. Plot of (B) ${\rm ln(\alpha)}$ with ${\rm
h\nu}$ for the same sample. The inverse of the slope gives the Urbach energy
of the sample.}
\label{egVthickness}
\vskip -0.3cm
\end{figure}

The band-gaps of the as-grown films were evaluated from the absorption 
spectrum using the standard Tauc method \cite{tauc}. Band-gap can be estimated 
by extrapolating the linear part of the plot between 
${(\rm \alpha h \nu)^{n}}$ and ${\rm h \nu}$, where `${\rm \alpha}$' is the 
absorption coefficient and ${\rm h \nu}$ has its usual meaning. The variable 
`n' takes the value of 2 or 0.5 for direct or indirect band-gap material, 
respectively. Although SnS is reported to have both indirect and direct 
band-gaps, we obtained linear fits for n=2 (fig~\ref{egVthickness}A), 
suggesting that our p-SnS as-grown films without exception had allowed 
direct band-gap~\cite{direct}. The band-gaps obtained for our samples are 
listed in Table I. It may be noticed that there is no or minor variation 
in the band-gap for the samples under study.

\begin{figure}[h!!!]
\begin{center}
\epsfig{file=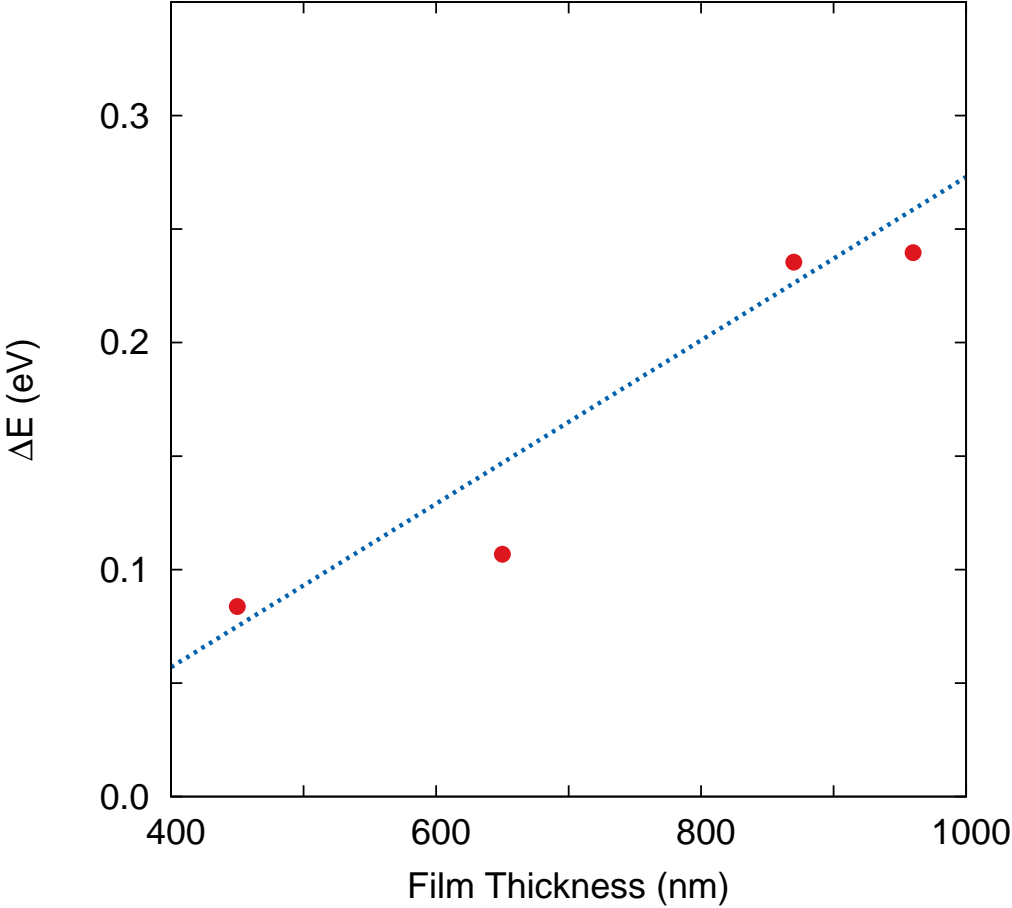, width=5.25in, angle=-90}
\end{center}
\caption{Variation of Urbach energy with film thickness.}
\label{Urbach}
\vskip -0.3cm
\end{figure}

The exponential region of the absorption spectra gives information on the 
localized states that tail off from the band edge within the band-gap,
usually called as Urbach energy (${\rm \Delta E}$). It arises due to
structural defects in the crystalline material~\cite{irib,arag}. In this 
region, 
the absorption coefficient, `${\rm \alpha}$', is given as
\begin{eqnarray}
\alpha=\alpha_oexp\left({h\nu \over \Delta E}\right)\nonumber
\end{eqnarray}
${\rm \Delta E}$ can be evaluated from the plot between ${\rm ln(\alpha)}$ and 
${\rm ln(h\nu)}$ (fig~\ref{egVthickness}B). It was found that ${\rm \Delta E}$ 
increased linearly with 
film thickness (fig~\ref{Urbach}). The significance of this result would be 
discussed in the following section.

\subsubsection{Refractive Index}
\begin{figure}[h!!]
\begin{center}
\epsfig{file=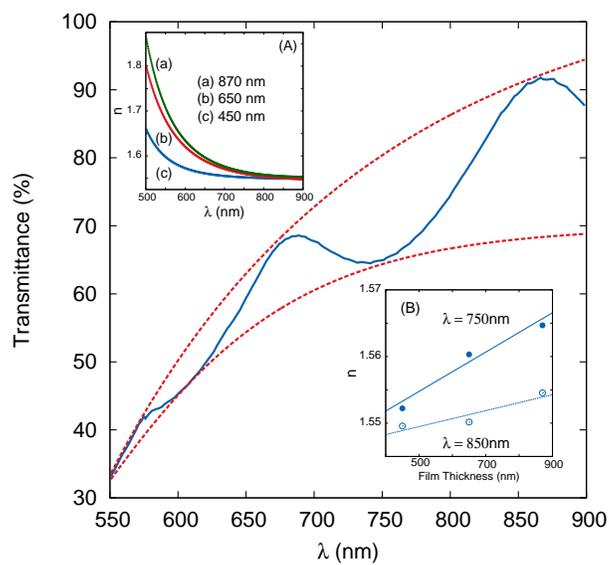, width=3.0in, angle=-90}
\end{center}
\caption{Transmission spectra of 450~nm thick film used to calculate 
refractive index by Swanepoel's method. Dotted line shows the envelopes
drawn joining the fringe maximas and minimas. Inset (A) shows the fit of
Sellmeier's model to the calculated refractive indices. Inset (B) shows the
linear variation in refractive index with film thickness for wavelengths 750
and 850~nm.}
\label{RefInd}
\end{figure}

Refractive Index of a thin film can be obtained from their transmission 
spectrum using the standard Swanepoel's method \cite{swan1,swan2}. 
Swanepoel's method involves drawing envelopes (fig~\ref{RefInd}) connecting 
the extreme 
points of the interference fringes appearing in the spectrum. Refractive Index 
in the transparent region is given as:
\begin{eqnarray}
n={\sqrt{N_1 + \sqrt{(N_1^{2} - s^{2})}}}\label{swan1}
\end{eqnarray}
\begin{eqnarray}
N_1={ {2s \over t_m} + {{s^{2} +1} \over 2}}\label{swan2}
\end{eqnarray}
where, n is the refractive index of the film, ${\rm t_m}$ is the minima of 
the interference fringes and `s' is the refractive index of the substrate  
(which in our case is glass and is taken as 1.5).

Interestingly, the fringes did not appear in very thin (thickness less than 
450~nm) and very thick samples (thickness greater than 870~nm), giving
context to our thickness range selection. In thicker samples, it seemed as
if the fringes were moving to higher wavelengths (${\rm >1000~nm}$). This was
observed by Yue et al~\cite{yue2} too. Since Swanepoel's method can not be 
applied in the spectra where fringes are absent, it restricts the film 
thickness on which calculations can be done and this presents as a major 
drawback of the method. In such cases, other methods like ellipsometry 
have to be employed for estimating the  refractive index. Due to lack of 
fringes in thicker samples and limitations imposed by the technique, we 
report the variation of refractive indices seen in just three of our samples, 
namely 450, 650 and 870~nm thick films (inset `A' of fig~\ref{RefInd}).

The refractive indices of SnS thin films showed normal dispersion relation 
i.e. it decreased with increasing wavelength. In fact the trend followed the 
Sellmeier relation \cite{fuji} (curve fits have co-relation of 0.998)
\begin{eqnarray}
n^2=A+\sum_j{B_j\lambda^2 \over \lambda^2-C_{oj}}\label{sell}
\end{eqnarray} 
It should be noted that while our data clearly showed that the refractive 
index of p-SnS films fit Sellmeier's model, previous works have reported that 
the SnS follows Cauchy's dispersion relation~\cite{botao} and 
Wemple-DiDomenico single oscillator model (WDD) for refractive 
index~\cite{wdd}.
Sellmeier model pictures each interband optical transistion as individual 
dipole oscillators such that one oscillator dominates and all the other
oscillators are combined together into and represented by the coefficient
`A'~\cite{wdd1}. 
Table~II gives the 
coefficients of eqn~(\ref{sell}) that fit to the experimental results. 
An increasing value of coefficient `B' with film thickness suggested that the 
refractive index of the samples increased with film thickness for all 
wavelengths which is validated by inset `B' of fig~\ref{RefInd}, which shows 
an increase in `n' values with film thickness for two wavelengths, 750 and 
850~nm.

Sellmeier model gives an empirical formula and fails to give an insight about 
the physical/ structural properties of the film. WDD model is an improved
dispersion model as it relates the optical properties with internal
structure by single electron oscillator approximation~\cite{yang}. WDD model 
is represented by the following equation~\cite{wdd2} 
\begin{eqnarray}
n^2=1+{E_dE_o \over E_o^2-(h\nu)^2}\label{wdd1}
\end{eqnarray} 
The WDD model gives a physical interpretation about the sample through these 
constants ${\rm E_o}$ and ${\rm E_d}$, where `${\rm E_o}$' is the average 
band-gap parameter also known as the 
oscillator energy and is proportional to the material's band-gap 
(${\rm \approx E_g}$) and `${\rm E_d}$' is the dispersion energy. ${\rm
E_d}$ is a measure of inter-band oscillation strength and is given as
\begin{eqnarray}
E_d=\beta N_cZ_aN_e\label{wdd2}
\end{eqnarray}
where `${\rm \beta}$' depends on the type of bond within the material
(${\rm \beta=0.26~eV}$ for ionic bonding and 0.37~eV for co-valent
bondings). `${\rm N_c}$' is the coordination number or the number of nearest 
neighboring cations, `${\rm Z_a}$' is the anion valency, while `${\rm
N_e}$' is the effective number of valence electrons per anion. We used 
the refractive index data obtained by Swanepoel's method above and found that 
the data did not fit into WDD model equation given by
eqn~(\ref{wdd1}). Considering that the refractive index values strongly 
depend on the drawn envelopes in Swanpoel method, one should not be surprised by 
the inconsistent results.

Refractive index of films can also be determined by ellipsometry. 
Banai et al~\cite{banai} have highlighted that spectroscopic ellipsometry
(SE) combined with UV-visible spectroscopy can yield more accurate results for 
optical properties. Also, since only a few literature is present~\cite{shaaban} 
on ellipsometry studies of SnS due to the difficult data analysis involved,
we decided to augment our results by carrying out SE studies on our SnS
samples.

\begin{figure}[h!!]
\begin{center}
\epsfig{file=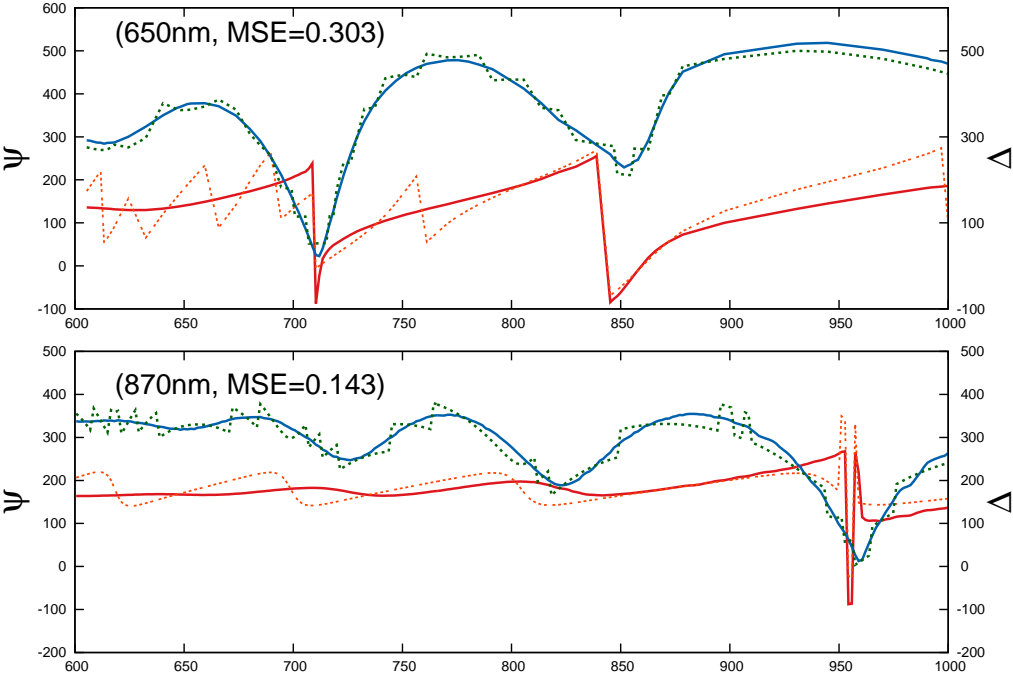, width=3.75in, angle=-90}
\end{center}
\caption{The experimental (solid line) and theoretically modelled (dotted
line) spectra for the two ellipsometric parameters (${\rm \psi}$ and ${\rm
\Delta}$) for film thickness of
650 and 870~nm.}
\label{SE}
\end{figure}

\begin{figure}[h!!]
\begin{center}
\epsfig{file=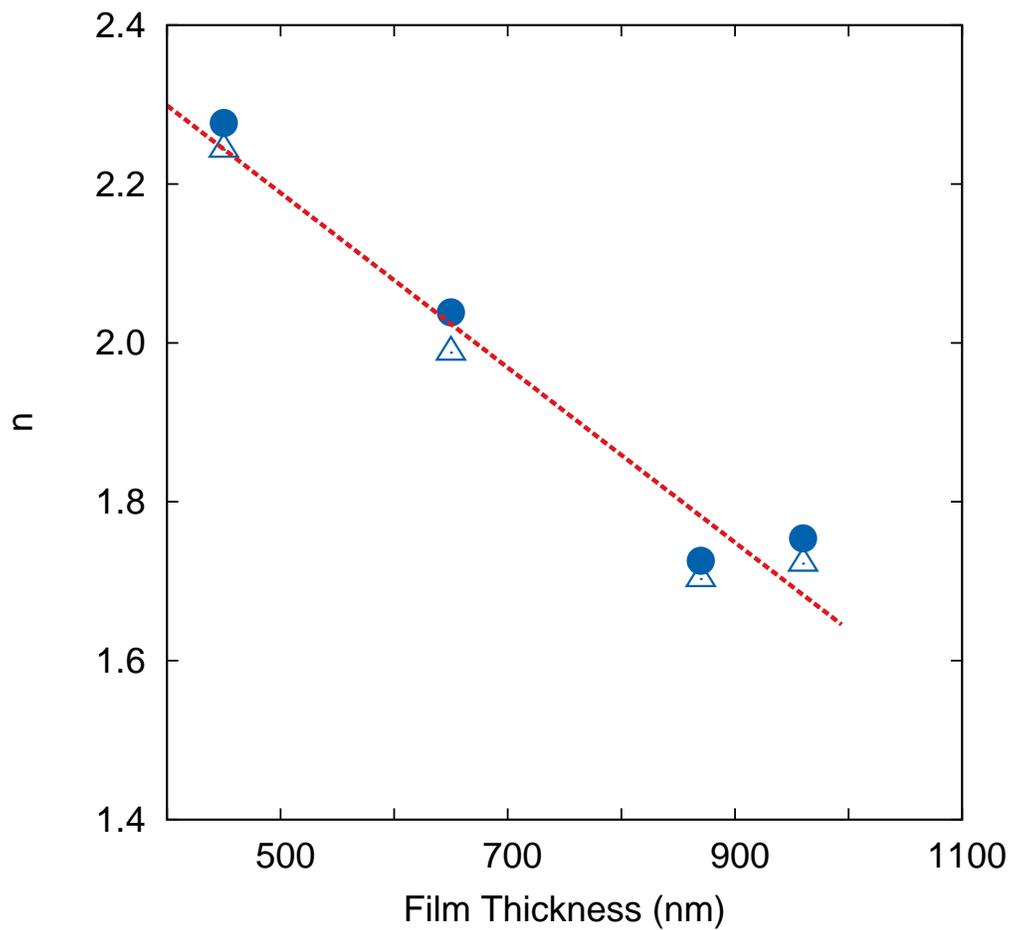, width=5.25in, angle=-90}
\end{center}
\caption{Variation of refractive index obtained from SE with thickness for
750~nm (circles) and 850~nm (triangles) wavelengths.}
\label{RefSE}
\end{figure}

\begin{figure}[h!!]
\begin{center}
\epsfig{file=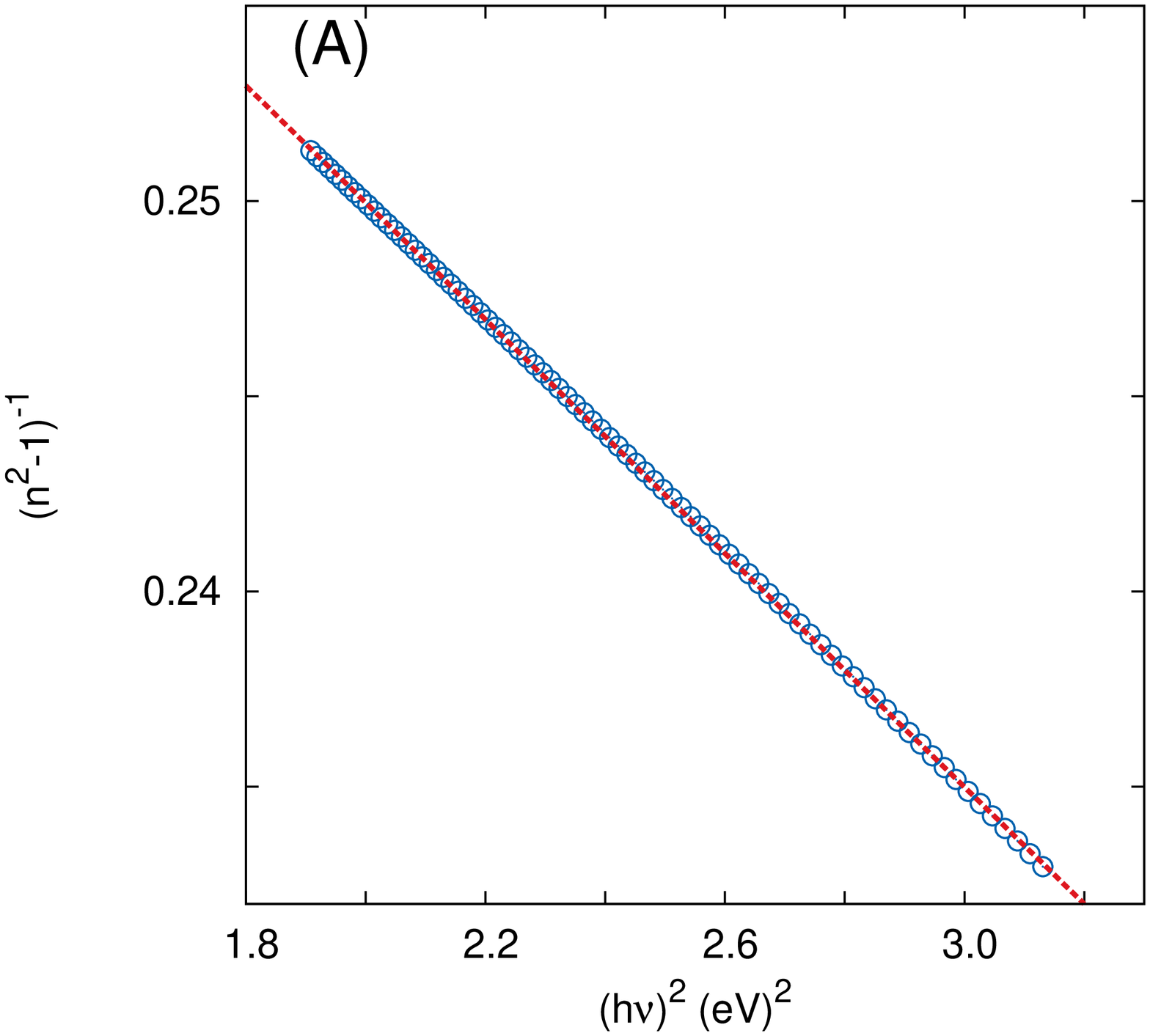, width=2.25in, angle=-90}
\hfil
\epsfig{file=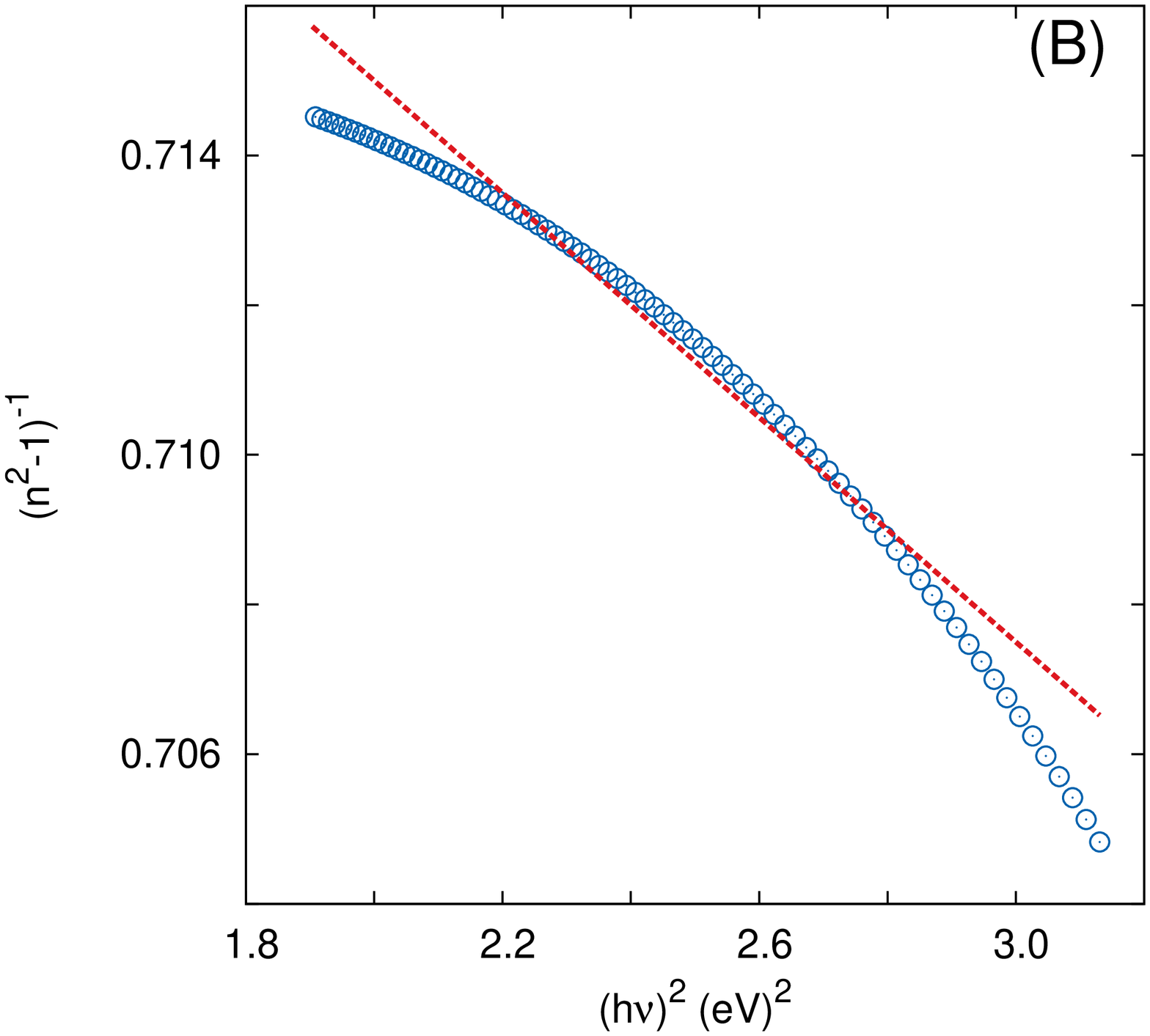, width=2.25in, angle=-90}
\end{center}
\caption{WDD model fitting for refractive index data obtained using (A)
Ellipsometry data, (B) Swanepoel's method. Slope (m) and intercept (c) were 
used to calculate ${\rm E_o (=\sqrt{c/m})}$ and ${\rm E_d (=1/sqrt{mc})}$.}
\label{wdd}
\end{figure}

The SE studies were done on our SnS films. Data of ${\rm \psi}$ and 
${\rm \Delta}$ collected were fit using two layer model, i.e. of film with
finite thickness on semi-infinite glass substrate. The optical 
parameters, n and k, of the glass substrate used in these calculations were 
also evaluated using SE data. The problem of back scattering was taken care of
by roughening the lower surface of glass substrate. The refractive index of
the glass substrate was found to follow the Cauchy's dispersion
relation~\cite{yash1}
\begin{eqnarray}
n &=& 1.489+{3.23 \times 10^3 \over \lambda^2}\nonumber\\
k &=& 6.89\times 10^{-4}+0.012exp\left[3.05\left({1240 \over \lambda}
-4.342\right)\right]\nonumber
\end{eqnarray}
The program iterated
different values of Sellmeier model's constants and film thickness for the 
SnS layer till a small value of Mean Square Error (MSE) was obtained showing
good convergence between fitted and experimentally obtained data 
(see fig~\ref{SE}). The constants of Sellmeier model obtained by SE data 
analysis are reported in Table~III. As can be seen from fig~(\ref{SE}), the 
data are oscillatory for ${\rm \lambda \leq 700~nm}$ and a good fit was not 
obtained, indicating that refractive index did not follow the Sellmeier 
dispersion model below 700~nm. Since we are interested in the transparent 
region, we had restricted ourselves for wavelength region 
${\rm 700<\lambda <800~nm}$ (i.e. the region where Sellmeier applies). Unlike 
the refractive index data obtained by Swanepoel's method, the refractive index 
obtained using SE was found to be decreasing with thickness. Fig~(\ref{RefSE}) 
shows the variation of `n' with film thickness for 750 and 850~nm wavelengths, 
a single straight line fitted for both the wavelengths due to minor variations 
in refractive index values at higher wavelengths where the Sellmeier model 
flats out. Also, these refractive index values were found to adhere to the WDD 
trend given by eqn~(\ref{wdd1}). The perfect linear fit obtained with the
SE data (see fig~\ref{wdd}A) was an improvement over the results obtained 
using Swanepoel method, which failed to fit the linear trend (see 
fig~\ref{wdd}B).

The prefect linear trend highlighted the accuracy of SE over Swanepoel method.
The improvement in refractive index data obtained from SE over
Swanepoel's can be understood considering there are often discrepancies
while deciding the strong transmission, weak-absorbing and strong absorbing
region of the film's absorption spectra on whose basis different
formulae have to be applied. Also, `n' is obtained by drawing envelopes for 
transmission graphs in the given wavelength range while in SE, `n' is
obtained by data fitting the model on two different variables (${\rm \psi}$
and ${\rm \Delta}$). Thus, as the number of data fitting variable increases,
the accuracy of the obtained refractive index values improves. 

Table~IV reports the values of ${\rm E_o}$ and ${\rm E_d}$ (eqn~\ref{wdd1}) 
evaluated from the graph (fig~\ref{wdd}A). While the values of ${\rm E_o}$ 
matched well with those of ${\rm E_g}$ obtained using Tauc's plot (literature 
reports ${\rm E_o \approx 2E_g}$)~\cite{selim}, there was a stark variation in 
the values of ${\rm E_d}$ as the film thickness varied. This was surprising 
considering that the XRD analysis and energy band-gap values suggested that 
the structure of the films were the same. However, the Urbach tail analysis 
suggested the presence of defects in the films, with ${\rm \Delta E}$ 
increasing with the film thickness. We believe the variation in ${\rm E_d}$ 
might be related to the defects in the films. 

\subsubsection{Photoluminescence}
\begin{figure}[h!!]
\begin{center}
\epsfig{file=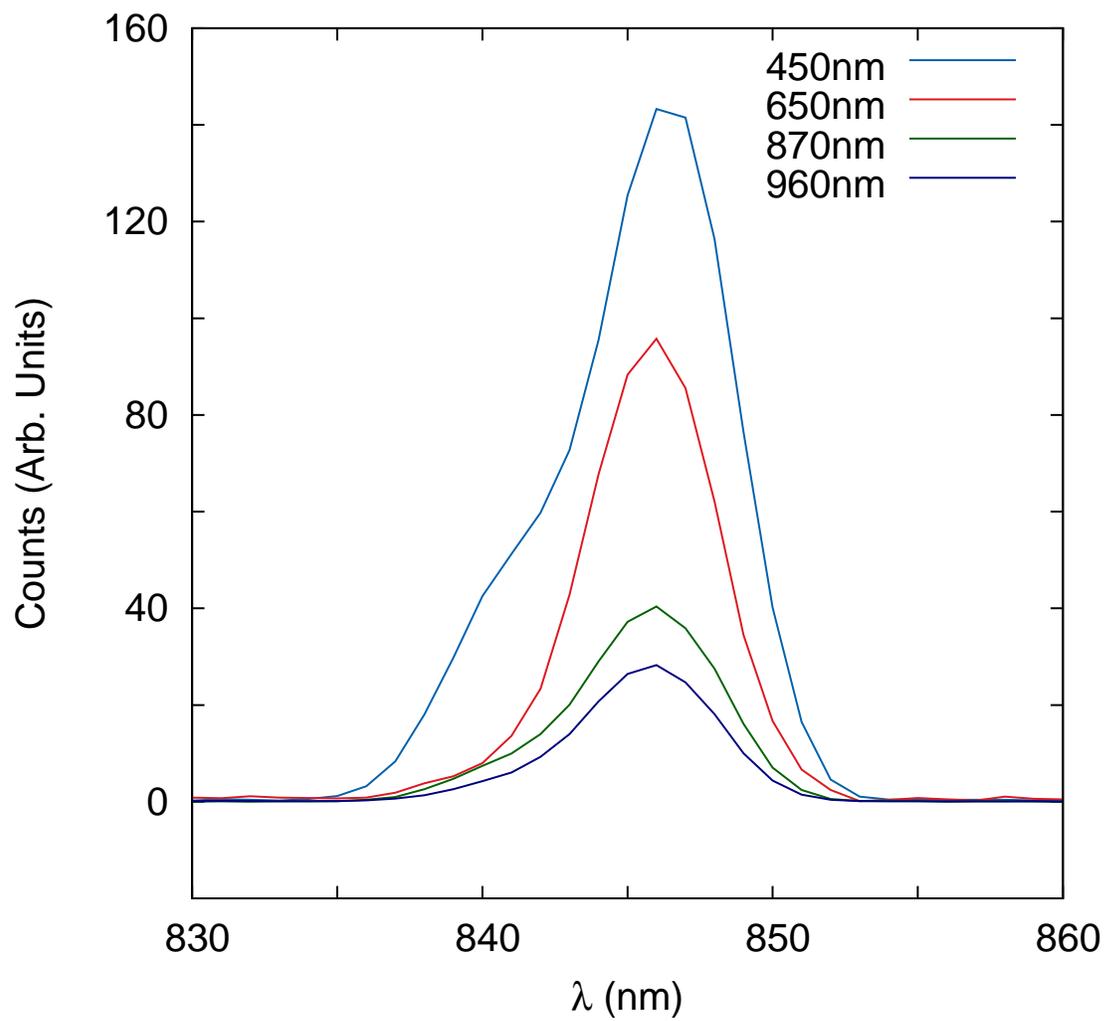, width=5.75in, angle=-90}
\end{center}
\caption{PL spectra for the asgrown films.}
\label{pl}
\end{figure}

To investigate the presence of defect levels within the band-gap, PL
measurements were made with the excitation wavelength of 550~nm.
Fig~(\ref{pl}) shows the observed PL spectra for all the films indicating a
broad peak around (830-860~nm) which can be deconvoluted into two peaks.
Since the band-gap of all the films was around 1.8~eV, these peaks could safely
be associated with the radiative transitions from/to defect levels within
the band-gap. Also, the presence of energy levels due to sulphur and tin
vacancies in SnS are well documented and was also linked to the persistent
photocurrent decay measurements of the as grown films~\cite{yppc}.

\begin{figure}[h!!]
\begin{center}
\epsfig{file=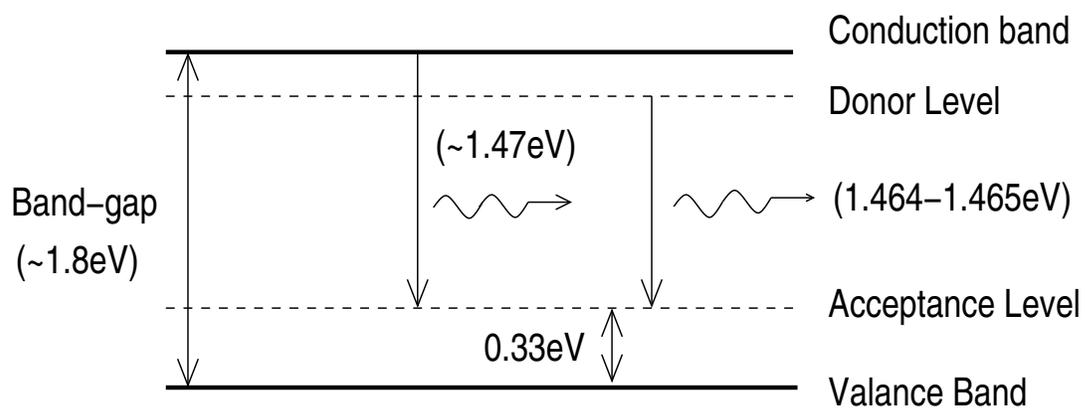, width=5.75in, angle=-0}
\end{center}
\caption{Basic energy level diagram of SnS showing the two possible
radiative transitions corresponding to the peaks in PL spectra.}
\label{elevel}
\end{figure}

\begin{figure}[h!!]
\begin{center}
\epsfig{file=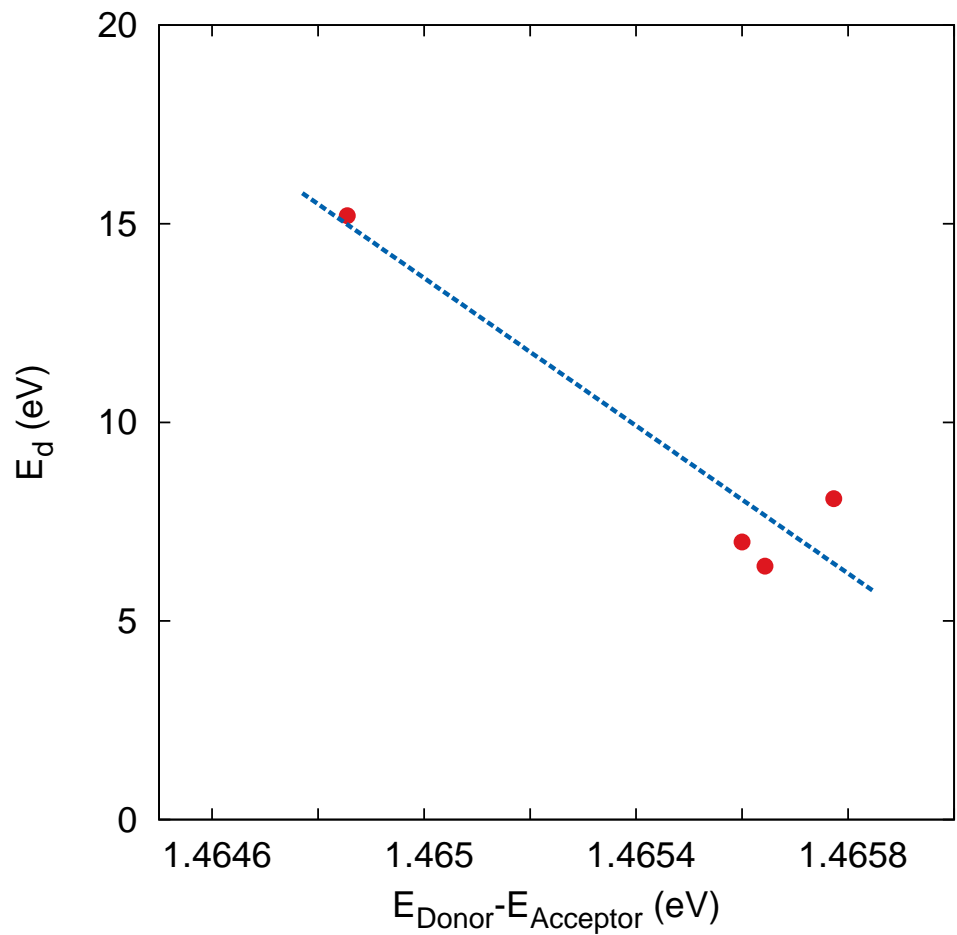, width=5.25in, angle=-90}
\end{center}
\caption{Variation in dispersion energy (${\rm E_d}$) with the energy
difference between donor and acceptor levels (calculated using PL data).}
\label{ellipsopaper}
\end{figure}

\begin{figure}[h!!]
\begin{center}
\epsfig{file=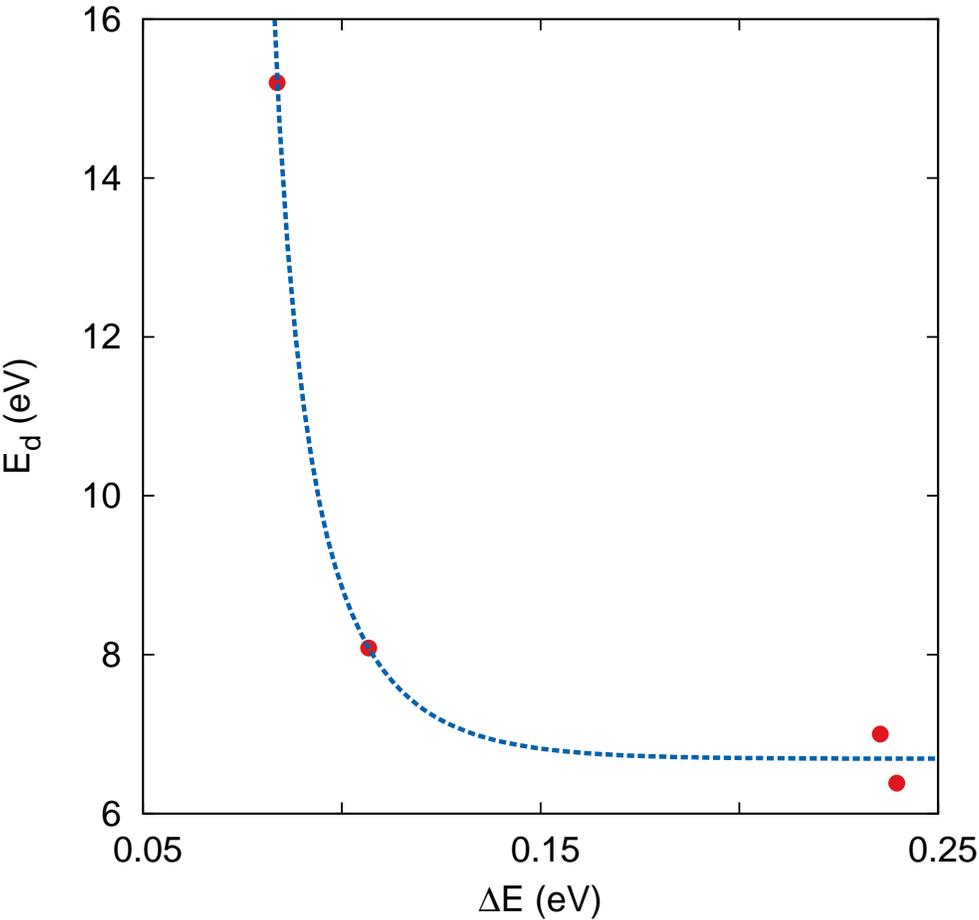, width=5.25in, angle=-90}
\end{center}
\caption{Variation in dispersion energy (${\rm E_d}$) with Urbach energy
(${\rm \Delta E}$).}
\label{Urbachfall}
\end{figure}

Fig~(\ref{elevel}) gives a crude schematic energy band level diagram of the
as grown SnS films. The two PL peaks correspond to the conduction
band (CB) to acceptor level transition and donor level to acceptor level
transition. Band-gap and donor to valence band (VB) transitions possibility
were ruled out due the lack of peaks in PL corresponding to the expected
energy levels. On co-relating our PL analysis with ${\rm E_d}$ values, we
find that ${\rm E_d}$ increased as the energy difference between the donor
and acceptor levels decreased (see fig~\ref{ellipsopaper}). Physically this
can be understood as the oscillation strength for the transition between the
levels increases as the difference in the energy levels decreases. A similar
trend was also seen between ${\rm E_d}$ and ${\rm \Delta E}$ of the Urbach
tail (fig~\ref{Urbachfall}) which confirmed the intimate relation of ${\rm
E_d}$ with defect structures and energy levels introduced by it. Increase in
${\rm \Delta E}$ with thickness indicated the increase in tailing or spread
of the band edges (increase in width of the localized states within the
band-gap) with thickness. Thus, validating our persistent photocurrent
results~\cite{yppc} where thinner samples showed a single exponential decay
curve due to small width of levels while as the width increased, multiple
transition levels appeared and a non-exponential decay curve was observed.

\section{Conclusions}
Nano-crystalline p-type SnS thin films grown on microscopy glass slide by 
thermal evaporation at room temperature were studied. Their optical properties 
were found to be dependent on the film thickness in the range 450-1000~nm. The
optical studies showed a band-gap of ${\rm \approx 1.8~eV}$ for the as-grown
films with the width of localized states (Urbach's tail) increasing with
thickness. Swanepoel's method was used for evaluating refractive index for
films using interference fringes in their transmission spectra. Refractive
index followed Sellmeier dispersion relation. The fitting coefficients were
used as initial guess for ellipsometric studies of the as grown films. The
improved dispersion data were validated by fitting WDD model, which indicated
that the defect levels affect the refractive index of the film. The study
also validated the PPC exponential decay for thinner and non-exponential
decay for our thicker samples. Thus, the desired optical properties of the
material can be obtained by material manuplication like changing thickness,
grain size or introducing defect levels for use in various applications.

\section{Acknowledgments}
One of the authors (YG) acknowledges DST~(India) for the financial support
extended under the INSPIRE program (Fellowship No. IF131164).

\clearpage
\section*{Figures}

\clearpage
\clearpage
\clearpage

\clearpage

\clearpage

\clearpage

\clearpage

\clearpage

\clearpage

\clearpage
\section*{Figure Captions}
\begin{itemize}
\item[1] X-ray diffractogram of film with thickness 960~nm.
Plot showing deconvolution of broad peak at 
${\rm 2\theta \approx 31^{\circ}}$ indicating (113) as the preferred 
orientation for 960~nm thick film.
\item[2] Plot shows variation in average grain size with thickness for the 
as-grown films. Inset shows spherical grains seen by Scanning Electron
Microscope. The micrograph exhibited here is of 870~nm thick SnS films.
\item[3] Plot exhibits variation of (A) ${\rm (\alpha h \nu)^2}$ with 
${\rm h\nu}$ for 650~nm thick film. Extrapolation of the best fit line to 
the X-axis at y=0 gives the band gap of the samples. Plot of 
(B) ${\rm ln(\alpha)}$ with ${\rm h\nu}$ for the same sample. The inverse of 
the slope gives the Urbach energy of the sample.
\item[4] Variation of Urbach energy with film thickness.
\item[5.]
Transmission spectra of 450~nm thick film used to calculate 
refractive index by Swanepoel's method. Dotted line shows the envelopes
drawn joining the fringe maximas and minimas. Inset (A) shows the fit of
Sellmeier's model to the calculated refractive indices. Inset (B) shows the
linear variation in refractive index with film thickness for wavelengths 750
and 850~nm.
\item[6.] The experimental (solid line) and theoretically modelled (dotted
line) spectra for the two ellipsometric parameters (${\rm \psi}$ and ${\rm
\Delta}$) for film thickness of 650 and 870~nm.
\item[7.] Variation of refractive index obtained from SE with thickness for
750~nm (circles) and 850~nm (triangles) wavelengths.
\item[8.] WDD model fitting for refractive index data obtained using (A)
Ellipsometry data, (B) Swanepoel's method. Slope (m) and intercept (c) were 
used to calculate ${\rm E_o (=\sqrt{c/m})}$ and ${\rm E_d (=1/sqrt{mc})}$.
\item[9.] PL spectra for the asgrown films.
\item[10.] Basic energy level diagram of SnS showing the two possible
radiative transitions corresponding to the peaks in PL spectra.
\item[11.] Variation in dispersion energy (${\rm E_d}$) with the energy
difference between donor and acceptor levels (calculated using PL data).
\item[12.] Variation in dispersion energy (${\rm E_d}$) with Urbach energy
(${\rm \Delta E}$).
\end{itemize}

\clearpage
\begin{table}[t]
{\bf Table I:} {\sl Band-gap values obtained for different film thicknesses using Tauc's method.}
\begin{center}
{\scriptsize {
\begin{tabular}{c c }
\hline 
  Film Thickness (nm) & Band-gap (eV)\\ \hline
  450   & 1.83\\
  650   & 1.80\\
  870   & 1.79\\
  960   & 1.84\\
\hline
\end{tabular}
}}
\end{center} 
\end{table}

\clearpage
\begin{table}[t]
{\bf Table II:} {\sl Constants of the Sellmeier model obtained via curve fit 
for different film thicknesses.}
\begin{center}
{\scriptsize {
\begin{tabular}{c c c c}
\hline 
  Film Thickness (nm) & A &  ${\rm B_1}$ & ${\rm C_{o1}}$ 
(${\rm \times 10^5}$)
${\rm nm^2}$\\ \hline
  450   & 2.278 & 0.0834 & 2.08\\
  650   & 1.983 & 0.3006 & 1.92\\
  870   & 1.963 & 0.3180 & 1.99\\
\hline
\end{tabular}
}}
\end{center} 
\end{table}

\clearpage
\begin{table}[t]
{\bf Table III:} {\sl Constants of the Sellmeier model obtained via SE data
analysis for different film thicknesses.}
\begin{center}
{\scriptsize {
\begin{tabular}{c c c c}
\hline 
  Film Thickness (nm) & A &  ${\rm B_1}$ & ${\rm C_{o1}}$ 
(${\rm \times 10^5}$)
${\rm nm^2}$\\ \hline
  450   & 2.2 & 2.4 & 1,1\\
  650   & 2.4 & 1.1 & 2.1\\
  870   & 2.0 & 0.7 & 1.6\\
  960   & 1.9 & 0.8 & 1.8\\
\hline
\end{tabular}
}}
\end{center} 
\end{table}

\clearpage
\begin{table}[t]
{\bf Table IV:} {\sl ${\rm E_o}$ and ${\rm E_d}$ obtained for various
thicknesses.}
\begin{center}
{\scriptsize {
\begin{tabular}{c c c}
\hline 
  Film Thickness (nm) & ${\rm E_o}$ &  ${\rm E_d}$ \\ \hline
  450   & 4.32 & 15.20 \\
  650   & 3.37 & 8.08 \\
  870   & 4.19 & 6.99 \\
  960   & 3.81 & 6.38 \\
\hline
\end{tabular}
}}
\end{center} 
\end{table}

\end{document}